# Fracture behavior of MOF monoliths revealed by nanoindentation and nanoscratch

Michele Tricarico and Jin-Chong Tan[*]

Multifunctional Materials and Composites (MMC) Laboratory, Department of Engineering Science, University of Oxford, Parks Road, Oxford, OX1 3PJ, United Kingdom.

**Email:** jin-chong.tan@eng.ox.ac.uk

## Abstract

Monolithic metal-organic frameworks (MOFs) represent a promising solution for the industrial implementation of this emerging class of multifunctional materials, due to their structural stability. When compared to MOF powders, monoliths exhibit other intriguing properties like hierarchical porosity, that significantly improves volumetric adsorption capacity. The mechanical characterization of MOF monoliths plays a pivotal role in their industrial expansion, but so far, several key aspects remain unclear. In particular, the fracture behavior of MOF monoliths has not been explored. In this work, we studied the initiation and propagation of cracks in four prototypical MOF monoliths, namely ZIF-8, HKUST-1, MIL-68 and MOF-808. We observed that shear faults inside the contact area represent the main failure mechanism of MOF monoliths and are the source of radial cracks. MIL-68 and MOF-808 showed a remarkably high resistance to cracking, which can be ascribed to their consolidated nanostructure.

## Significance Statement

Fabrication of metal-organic frameworks (MOFs) as bulk shapes plays a key role in the translation of these functional materials from laboratories to real-world applications. Particularly, sol-gel derived MOF monoliths are promising candidates, due to their combination of structural stability and hierarchical porosity, which allows for increased gas uptake. The mechanical response of monolithic MOFs is poorly understood and, especially the fracture phenomenon underpinning polycrystalline monoliths has not yet been elucidated. Here we investigate the cracking behavior of monoliths under indentation: we observed a remarkable resistance to radial cracking and the occurrence of shear faults as the principal failure mechanism. Exact knowledge of structure-property relationships will guide the design of mechanically tough MOF monoliths, with an excellent resilience for practical use.

## Introduction

Metal-organic frameworks (MOFs) are hybrid materials composed of metal nodes and organic linkers that self-assemble to form one-, two-, or three-dimensional lattice structures (1, 2), characterized by a large internal surface area, which makes MOF suitable for a wide range of applications (3-6).

However, nowadays industrial applications are limited by the morphology of this class of materials, which are mostly available as polydisperse microcrystalline powders. A way to overcome this limitation is the realization of "monolithic" materials, which, together with bulk morphology and structural stability, offer other advantages such as reduced mass transfer resistance (for improved gas separation capabilities) and higher volumetric adsorption capacities compared to the powder counterparts (7). One of the most frequently used techniques to synthetize MOF monoliths is the sol-gel processing route (8-12). Sol-gel monoliths are formed by removing solvent from a gel by slow drying, generating a polycrystalline bulk solid. This method allows also to obtain different pore sizes within the same material, forming hierarchical structures which are desirable for applications (13).

The mechanical properties of MOF monoliths (and MOF in general (14)) have been measured mostly by nanoindentation, which allows to probe very small volume of material and yet providing accurate measurements of Young's modulus ($E$) and hardness ($H$). In our previous study (11), we explored the correlation between the nanostructure and the mechanical response of two prototypical MOF monoliths, namely ZIF-8 and ZIF-71, by employing nanoindentation, atomic force microscopy (AFM) and nearfield infrared (IR) nanospectroscopy, along with finite element simulations. We were able to identify grain boundary sliding (GBS) as the main deformation mechanism in crystalline MOF monoliths, followed by densification caused by framework collapse for higher pressures. At the end of the paper, we also reported an estimation of the fracture toughness ($K_{IC}$) of the two monoliths, by measuring the length of radial cracks induced by cube corner indenter. Yet, the cracking phenomenon is not understood — for a general class of polycrystalline MOF monolithic materials.

The aim of this work is to focus on the nanoindentation fracture behavior of MOF monoliths, that, to the best of our knowledge has never been systematically studied to reveal the underpinning mechanisms. In fact, only a few studies have explored the field of MOF fracture, but these are limited to specific single crystals (15) and glasses (16, 17).

Nanoindentation techniques have been widely used to evaluate the fracture toughness of brittle materials, using sharp tip geometries such as Vickers or cube corner. These measurements rely on empirical models that correlate $K_{IC}$ with the length of the radial cracks originating from the corners of the indenter marks (18). However, initiation and the propagation of such cracks is not easy to model. Furthermore, cracking can occur also underneath the contact area, *i.e.* median and lateral cracks (19), and other parameters have been shown to complement toughness in the fracture mechanisms, such as the hardness-to-modulus ratio ($H/E$) (20). Although the fracture toughness values for two ZIF monoliths have been estimated in our recent study (11), the origin of the failure, particularly in connection to the nature of crack initiation and propagation was unclear.

In addition, nanoscratch tests were conducted for the first time on crystalline MOF monoliths. This type of test is a well-established technique to characterize the wear resistance of metals (21), ceramics (22) and polymers (23). Nanoscratch experiments have been used to study the cohesion and adhesion properties of MOF films (24, 25) and a MOF glass (26), but never in the context of the fracture of nanocrystalline MOF monoliths.

In this paper, we employed nanoindentation, microindentation, and nanoscratch on four prototypical MOF monoliths: ZIF-8, HKUST-1, MIL-68 and MOF-808. We have observed initiation and propagation of the cracks following a sharp contact load, and established a correlation between the fracture behavior and the elastic-plastic response of the materials and their underlying nanostructures.

**Results**

We identified four prototypical MOFs that form crystalline monoliths for the fracture studies (see Figure 1), namely ZIF-8, HKUST-1, MIL-68 and MOF-808, characterised by different structures:

- ZIF-8 (Zn(mIm)$_2$, mIm = 2-methylimidazolate) consists of the ZnN$_4$ tetrahedral metal centres (Zn(II)) coordinated by N atoms of the mIm linkers. An angle of 145° is formed at the Zn−mIm−Zn centre, mimicking the Si−O−Si angle in zeolites. Its sodalite architecture results in a pore size (diameter of the largest sphere that will fit into the nanocage) of 11.6 Å. Its chemical and thermal stability make this material suitable for gas adsorption and separation applications (27).
- HKUST-1 (Cu$_3$BTC$_2$, BTC = benzene-1,3,5-tricarboxylate) is a faced-centred cubic crystal, which contains an intersecting 3D system of square-shaped pores of 9 Å × 9 Å. The framework is built up of dimeric metal units (Cu(II)), which are connected by BTC ligands, forming a paddlewheel motif, also called secondary building unit (SBU) (28).
- MIL-68 is characterised by a 3D network with a Kagomé-like lattice, made of infinite trans-connected chains of octahedral units InO$_4$(OH)$_2$ linked to each other through terephthalate ligands, resulting in triangular and hexagonal one-dimensional channels (window diameters of 6 Å × 16 Å, respectively) (29).

- MOF-808 ($Zr_6O_4(OH)_4(BTC)_2(HCOO)_6$) is built up of zirconium oxide SBUs, consisting of six octahedrally coordinated zirconium atoms held together by eight $\mu_3$-oxygen atoms, each connected to six BTC linkers. The coordination of SBU is completed with six non-structural ligands, providing charge compensation. This framework architecture yields tetrahedral cages (pore diameter 4.8 Å) with SBUs on the vertices and linkers on the faces (tertiary building units, TBUs). The MOF-808 framework is finally obtained by the TBUs sharing corner, in such a way that a large adamantane-shaped pore is formed (pore diameter 18.4 Å) (30).

We measured the mechanical properties of the four monoliths by nanoindentation, as shown in Figure 1A and Figure S2. We used a Berkovich indenter tip to probe the indentation modulus ($E^*$) and hardness ($H$), see Materials and Methods for details. The above-mentioned mechanical properties are reported in Table S1. MIL-68 and HKUST-1 show the higher elastic modulus, with the latter also exhibiting an outstanding hardness, exceeding the one already reported for this material (9). The elastic energy was also computed on the basis of $W_{elastic}/W_{total}$ (see *Methods*), and plotted against the $H/E$ ratio in Figure 1B. In this plot we distinguish three zones, corresponding to the high, intermediate, and low elastic recovery (31, 32). It can be seen that, ZIF-8 and ZIF-71 monoliths reported by a previous work of the authors (11), and ZIF-62 glass reported by Stepniewska *et al.* (16), all fall into the category of high elastic recovery. HKUST-1 monolith shows an intermediate recovery, while the monoliths of MIL-68 and MOF-808 show a low elastic recovery. This wide difference in elastic recovery may be explained by the distinct architectures, in terms of framework topology (33) and chemical bonding (34, 35), among the four materials. The large pore size of MIL-68 and MOF-808 makes them more prone to deform permanently (lower $H$), while the high elastic recovery of ZIF-8 may be ascribed to its relatively smaller pore size and the lower coordination number of the zinc metal node, which governs the framework flexibility of the sodalite cage (lower $E$) (36).

In Table S1, we compared the residual depth at the end of the test (from the indentation load-depth curves) with the residual depth observed from AFM height topography of the residual indents (Figure S3). We noticed a large discrepancy in residual indentation depths, which demonstrates a viscoelastic recovery of the residual imprint over time (37). Such a time-dependent mechanical recovery was relatively limited for ZIF-8 and MIL-68 (~27% and ~34%, respectively), while it is significant for HKUST-1 and MOF-808 (67% and 43%, respectively).

With the aim of inducing radial cracks, we probed the monoliths with a cube-corner indenter (Figure S4), applying a load of 50 mN (maximum capacity of the nanoindenter load cell). Only ZIF-8 and HKUST-1 monoliths show crack propagation from the indent corners, while MIL-68 and MOF-808 monoliths are completely radial crack-free in the apparent vicinity of the residual indent. Subsequently, we estimated the fracture toughness of the HKUST-1 monolith by employing Laugier's empirical formula (18), adapted for a cube-corner indenter (38). We obtained $K_{IC}$ = 0.80 ± 0.45 $MPa\sqrt{m}$, which is the highest reported value so far for MOF materials. Notably, it is significantly higher than the monoliths of ZIF-8 (0.074 ± 0.023 $MPa\sqrt{m}$) and ZIF-71 (0.145 ± 0.050 $MPa\sqrt{m}$) reported in (11), and it is also considerably higher than previously reported ZIF-62 glass (0.104 ± 0.020 $MPa\sqrt{m}$) (17) and against dense hybrid frameworks single crystals ($K_{IC}$ ~ 0.1 – 0.3 $MPa\sqrt{m}$) (15).

In an attempt to crack also MIL-68 and MOF-808 monoliths, we used a microhardness Vickers indenter, which is capable of reaching much higher loads. Surprisingly, we found that none of the samples exhibited radial cracks from the indent corners, even with a load of 50 gf (~0.49 N). However, we observed the formation of layered shear faults in the contact area (Figure 2), similarly to what Stepniewska *et al.* (16) reported for ZIF-62 glass. The distance between two consecutive faults is different in all the materials, but it is much larger in ZIF-8. We observed cracks for even higher loads: 500 gf (~4.9 N) for MIL-68 and 300 gf (~2.9 N) for MOF-808, as shown in Figures S5 and S7, respectively. However, when we repeated the test with the same parameters on MIL-68, no radial crack was observed (Figure S6). The radial cracks observed in Figures S5 and S7 initiate from the shear faults inside the indents and propagate in a catastrophic fashion. Upon propagation, the cracks deflect following unpredictable paths; we reasoned that this occurs along the low-energy grain boundaries present in the MIL-68 and MOF-808 monoliths. Importantly, such cracking behavior found in a polycrystalline monolith is distinct from that observed for the amorphous ZIF-62 glass (no grain boundaries) which exhibits radial cracking from all four corners of Vickers indentation (16).

According to what we have reported in our previous work on ZIF monoliths (11), we observed a nanograined structure (Figure 3), with each nanograin being a single crystal. AFM phase images in Figure 3 and height topographies in Figure S9 reveal the morphology and the size of these nanograins. Indeed, we can clearly distinguish two different kinds of polycrystalline aggregates. For ZIF-8 and HKUST-1, which were synthesized by leveraging the high-concentration reaction (HCR) method (39), we observe a "nanoplate" morphology of the nanocrystals, with a maximum size of approximately 100 nm. Such nanoplates form aggregates by stacking on top of each other. In contrast, the synthesis of MIL-68 and MOF-808 resulted in relatively smaller nanograins, which tend to aggregate forming polycrystalline "lumps". The AFM height topographies shown in Figure S9 reveal that MIL-68 and MOF-808 nanograins are approximately 3 times smaller than the ZIF-8 and HKUST-1 counterparts. The above finding supports the notion that this particular nanostructural feature, which results in an increased volume fraction of grain boundaries, together with the framework structure, contributes to the outstanding ductility of MIL-68 and MOF-808 monoliths in compression; hence their correspondingly low $H/E$ ratios (Figure 1B).

Nanoscratch experiments were performed employing a Berkovich tip. Each test consists of three sequential steps: (i) a small load is applied allowing for the tip to track the pre-scratch surface profile; (ii) the same process is repeated during the actual scratch phase, with the prescribed applied load; (iii) the post-scratch profile and a cross-section profile (at half the length of the scratch) is recorded in order to measure the residual deformation after elastic recovery. We set the maximum load to 50 mN and a scratch length of 100 µm, the scratch velocity was set to 10 µm s$^{-1}$. The test can be performed in two distinct modes (Figure S10), depending on which end of the tip is cutting the material: ploughing (sharp end) or pushing (flat end). We observed systematically higher scratch critical depths in pushing mode compared to those in ploughing mode (Figure 4 and Table S2). This is due to the larger area of the flat end of the tip, with the same load, is able to remove a larger amount of material compared to the sharp counterpart. From the micrographs in Figure S12, we notice that only the ZIF-8 monolith shows evidence of crack events (outward from the scratch direction), while the other monoliths exhibit a very good ductility. Moreover, no pile-up is observed around the scratch (Figure S11).

Finally, we employed nearfield infrared nanospectroscopy (nanoFTIR) to gain further insights into the pressure-induced structural modifications of the framework within the monoliths. Local-scale IR absorption spectra, with a spatial resolution of ~20 nm, were taken inside the Vickers residual indent of ZIF-8 (the only sample that the AFM tip was able to probe thanks to its large elastic recovery). The point-to-point spectra taken on and around the shear faults (Figure 5A-B) are resembling the ones far away from the indented area, suggesting that the framework preserves its structural integrity. When we move the probe to the apex of the indent (Figure 5C-D), where the pressure is supposed to be the highest, we observed some changes in the characteristic peaks of ZIF-8. Let us consider the spectrum taken at the center of the indent (in green in Figure 5C): a new peak appears at 970 cm$^{-1}$ and the intensity of the 1124 cm$^{-1}$ band, relatively to the characteristic peak at 1145 cm$^{-1}$, increases. Also, the relative intensity of 1311 cm$^{-1}$ increases. According to Möslein *et al.* (40) the peak at 1311 cm$^{-1}$ can be assigned to the missing linker defect in the framework of ZIF-8. This result is in line with what the authors observed in a previous work (11) inside a Berkovich residual indent, which is characterized by an apex angle similar to the one of a Vickers indenter. The absence of defects in the vicinity of the shear faults suggests that the material fails along the grain boundaries. This process dissipates energy, relieving the stress applied to the grains, which therefore do not become amorphized.

**Discussion**

As illustrated in Figure 1B, the monoliths lying in the high-intermediate elastic recovery region are the only ones exhibiting radial cracks. Interestingly, also the ZIF-62 glass reported by (16), is located in the high elastic recovery region and evidently radial cracks are observed propagating from the shear faults inside the contact area.

The correlation between elastic recovery and radial cracks can be explained with the model for indentation fracture theory proposed by Lawn *et al.* (41) for ceramic materials. When a sharp indenter contacts and starts penetrating the surface, a plastically deformed zone develops about the indenter and a sub-surface tensile stress field is generated immediately beneath the tip. This leads to the formation of sub-surface median cracks. Upon unloading, the median cracks would want to close up, but are prevented in doing so by the existence of a residual stress field due to an elastic-plastic mismatch existing at the border of the plastic zone, caused by the material attempting to accommodate the plastically deformed zone. The residual stress is also the cause of lateral cracks (19). In their work on soda lime glasses, Lawn *et al.* (42) observed that radial cracks originate from the shear faults and then propagate upon unloading when the normal stress on such a crack is tensile. The indentation stress field can be divided into a reversible (elastic) and irreversible (residual) component (41). During loading, only median cracks, which give rise to shear faults, are formed, driven mainly by elastic components of the stress field. Lateral and radial cracks are formed upon unloading when the driving force is represented by residual stresses.

We propose that this mechanism, observed for glasses and other brittle materials, can apply to the MOF monoliths. The high elastic recovery of ZIF-8 monolith upon unloading provides the driving force for radial crack propagations, initiating from the shear faults and propagating towards the external surface as evidenced in this work.

The extraordinary resistance to radial cracking of MIL-68 and MOF-808 monoliths, which exhibit low elastic recovery, can be ascribed to their nanostructure. The reduced size of the nanograins and the consequently higher density of grain boundaries causes the material to undergo large "plastic" deformations by means of grain boundary sliding (GBS) in the vicinity of the contact area, thereby hindering the elastic propagation of the crack from the shear faults. The viscoelastic recovery of the residual imprints, a time-dependent effect, is also likely to play a role in preventing crack propagation by crack closure. Combined, these mechanisms help to enhance the toughness of the MIL-68 and MOF-808 monoliths.

From the scratch tests we observed how MIL-68 and MOF-808 monoliths are more ductile and they can undergo large plastic deformation, being able to accommodate large penetration depths (in both ploughing and pushing modes) without any sign of cracking. On the contrary, ZIF-8 is not able to sustain such a plastic deformation and dissipates energy by brittle fracture. We ascribe this behavior to the nanostructure of the monoliths: the higher volume fraction of grain boundaries in MIL-68 and MOF-808 promotes GBS and hence enabling plastic deformation.

Plastic flow is usually associated with material pile-up around the indenter; however, such a phenomenon is not observed in this case (Figure S11). We reasoned that a continuous flow of material is prevented by the stepwise shear-activated failure of the material, which, given the small size of the nanograins, it is well contained inside the scratch area and does not result in any catastrophic cracking or chipping events. This phenomenon was confirmed by local nanoFTIR spectra taken inside the Vickers residual imprint of ZIF-8 (Figure 5). In the vicinity of the shear faults, the material does not exhibit signs of pressure-induced structural amorphization, suggesting that the material tends to fail along the grain boundaries instead, dissipating energy.

In summary, the basic new insights derived from our research on the fracture and cracking behavior of nanocrystalline MOF monoliths will open the door to the nanostructural engineering and shaping of a class of *mechanically resilient framework solids* fit for real-world applications.

**Materials and Methods**

***Synthesis and sample preparation***

**ZIF-8 monoliths** ZIF-8 (Zn(mIm)$_2$) monoliths were synthesized following the procedure described in (11) : 0.595 g of Zn(NO$_3$)$_2$·6H$_2$O and 0.493 g of mIm were dissolved in 9 mL of DMF each and stirred for 5 minutes. Then, 0.837 mL of triethylamine (NEt$_3$) were added to the linker solution. Subsequently, the two solutions were combined in a 50 mL vial, where a gel was promptly formed. The molar ratio of

Zn(NO$_3$)$_2$·6H$_2$O : mIm : DMF : NEt$_3$ used was 1 : 3 : 116 : 3. The mixture was sonicated for 5 minutes and then washed three times, in 50 mL of solvent (DMF, MeOH and MeCN, respectively), followed by centrifugation at 8,000 RPM. The collected solid was dried slowly at room temperature (RT ~25 °C) for 3 days under the fume cupboard to yield monoliths.

**HKUST-1 (Cu$_3$BTC$_2$) monoliths** 300 mg of benzene-1,3,5-tricarboxylate (BTC) were dissolved in 10 mL of ethanol and subsequently 0.519 mL of NEt$_3$ were added to the solution. 270 mg of copper nitrate were dissolved in 10 mL of EtOH and added to the linker solution. The resulting solution was stirred for 15 minutes and washed three times in EtOH. A gel was collected and dried at RT for 2 days, yielding millimeter-sized "glassy" monoliths.

**MIL-68(In) monoliths** 9 mL DMF solution of 797 mg 1,4-benzenedicarboxylate (BDC) plus NEt$_3$ (9.6 mmol) were dissolved together. After that, 9 mL DMF solution of 1444 mg indium nitrate was immediately added into the mixture. Then the product was washed thoroughly 4 times (2 times with DMF, 2 times with MeOH). The nanocrystals of MIL-68(In) were separated from the suspension by centrifugation at 8,000 rpm for 10 mins and the excess solvent decanted. The obtained material was dried at room temperature for 3 days to yield monoliths.

**MOF-808 monoliths** 210 mg of benzene-1,3,5-tricarboxylate (BTC) and 970 mg of zirconyl chloride octahydrate were dissolved in a DMF/formic acid (30 mL + 30 mL), placed in a glass reagent bottle and heated at 130 °C for two days. The solution was centrifuged, and the collected material was washed four times with DMF. The resulting compound was then soaked in 100 mL of acetone for four days, with the solvent being replaced twice per day. The acetone-exchanged sample was finally evacuated at room temperature for 24 h and activated at 150 °C for 24 h to yield the monoliths.

Powder X-ray diffraction (XRD) patterns (Figures S1) confirm the successful synthesis of the MOF structures, since all the main characteristic Bragg diffraction peaks are present in the resultant monoliths.

The as-synthetized monoliths were cold-mounted in epoxy resin (Struers Epofix), resulting in a cylindrical sample, suitable for nanoindentation. In order to get reproducible results from the indentation tests, the contact surface must be flat. Therefore, the mounted specimen surface was carefully ground with different grades of emery papers, followed by fine polishing in diamond suspensions.

*Nanoindentation tests*

Indentation modulus ($E^*$) and hardness ($H$) were measured following the approach proposed by Oliver and Pharr (43):

$$E^* = \frac{\sqrt{\pi}}{2} \frac{S}{\sqrt{A(h_{\max})}} \quad (1)$$

$$H = \frac{P_{\max}}{A(h_{\max})} \quad (2)$$

where $S$ is the contact stiffness (slope of the unloading curve at maximum load), $A(h)$ is the area function, and $P_{\max}$ and $h_{\max}$ are the maximum load and depth, respectively. The indentation modulus ($E^*$) is a function of the Young's moduli and Poisson's ratios of the sample ($E_s$, $v_s$) and the indenter ($E_i$, $v_i$):

$$\frac{1}{E^*} = \frac{1 - v_s^2}{E_s} + \frac{1 - v_i^2}{E_i} \quad (3)$$

The area function $A(h)$ is a 3$^{rd}$ order polynomial that relates the contact area to the contact depth, and it is determined through calibration using a fused silica sample.

From the load-depth curves, we also computed the elastic recovery, $W_{elastic}/W_{total}$, defined as the ratio between the area under the unloading and loading curves respectively:

$$\frac{W_{\text{elastic}}}{W_{\text{total}}} = \frac{\int_{h_f}^{h_{\max}} P_{\text{unloading}}\, dh}{\int_0^{h_{\max}} P_{\text{loading}}\, dh} \qquad (5)$$

where $h_{max}$ is the maximum indentation depth, $h_f$ the residual depth after unloading, and $P_{unloading}$ and $P_{loading}$ are the loads applied upon loading and unloading, respectively.

We employed an iMicro nanoindenter (KLA-Tencor) equipped with a Berkovich tip. The Continuous Stiffness Measurement (CSM) method was employed to continuously measure the change in mechanical properties as a function of the indenter tip penetration depth. This technique superimposes a 2-nm oscillation on the quasi-static force, using a frequency-specific amplifier to measure the response of the indenter. The measurements were conducted by setting a maximum depth of 1,000 nm and an indentation strain rate of 0.2 $s^{-1}$. The values for $E$ and $H$ were computed by averaging the CSM data between 500 and 1000 nm.

For fracture studies, a cube corner indenter (three-sided pyramid with mutually perpendicular faces) was used to induce cracking, since its sharpness produces much higher stresses and strains in the region of contact compared with a Berkovich tip.

### *Vickers microindentation tests*

Vickers microindentation was performed using a Duramin-40 hardness tester (Struers). The maximum load (measured in gf) was held for 10 s before unloading.

### *Nanoscratch tests*

Nanoscratch tests were carried out on an iMicro nanoindenter (KLA-Tencor) equipped with a Berkovich tip. The maximum load was set to 50 mN, the scratch length to 100 μm, and the scratch velocity to 10 μm $s^{-1}$ for all the tests.

### *AFM imaging*

The surface topography of the monoliths and the drop-casted aggregates were measured by atomic force microscopy (AFM) as implemented in a neaSNOM instrument (neaspec GmbH) under the tapping-mode. A Scout350 (NuNano) probe was employed, with a nominal tip radius of 5 nm and a resonant frequency of 350 kHz.

### *NanoFTIR*

Nearfield Fourier transform infrared nanospectroscopy (nanoFTIR) was performed using the s-SNOM instrument (neaspec GmbH), where a platinum-coated AFM probe (Arrow-NCPt, tip radius < 25 nm, 285 kHz) under the tapping mode is illuminated by a broadband mid-infrared (IR) laser source (Toptica). Local nano-FTIR spectra of specific regions inside and away from the residual indentations were measured under a spot size of 20 nm. Each point spectrum was acquired as an average of 12 individual interferograms taken on the same spot, with 1024 pixels and an integration time of 14 ms per pixel, normalized by a reference spectrum taken on a silicon wafer.

**Acknowledgements**

We thank the ERC Consolidator Grant (PROMOFS grant agreement 771575) and EPSRC Impact Acceleration Account Award (EP/R511742/1) for funding the research. We are grateful to Dr. Igor Dyson (Oxford Engineering) for providing access to the microhardness Vickers indenter.

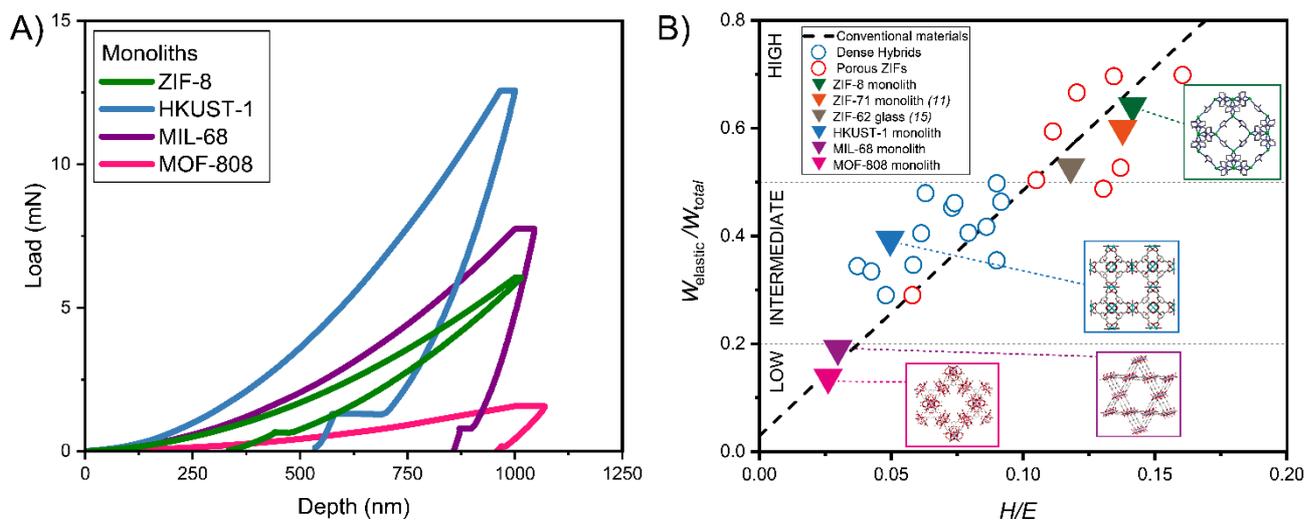

**Figure 1.** A) Representative load-depth curves resulting from nanoindentation of the monolithic samples of ZIF-8, HKUST-1, MIL-68 and MOF-808, using a Berkovich indenter. The tests were performed under displacement control, with the maximum surface penetration depth set to 1000 nm. The maximum load was held for 1 s to assess creep deformation; during unloading the load was held constant at 10% of the maximum load to quantify thermal drift (horizontal segment < 2 mN). B) Map of elastic recovery ($W_{elastic}/W_{total}$) vs the ($H/E$) ratio, the data for conventional materials, dense hybrid frameworks, and porous ZIFs were adapted from reference (32).

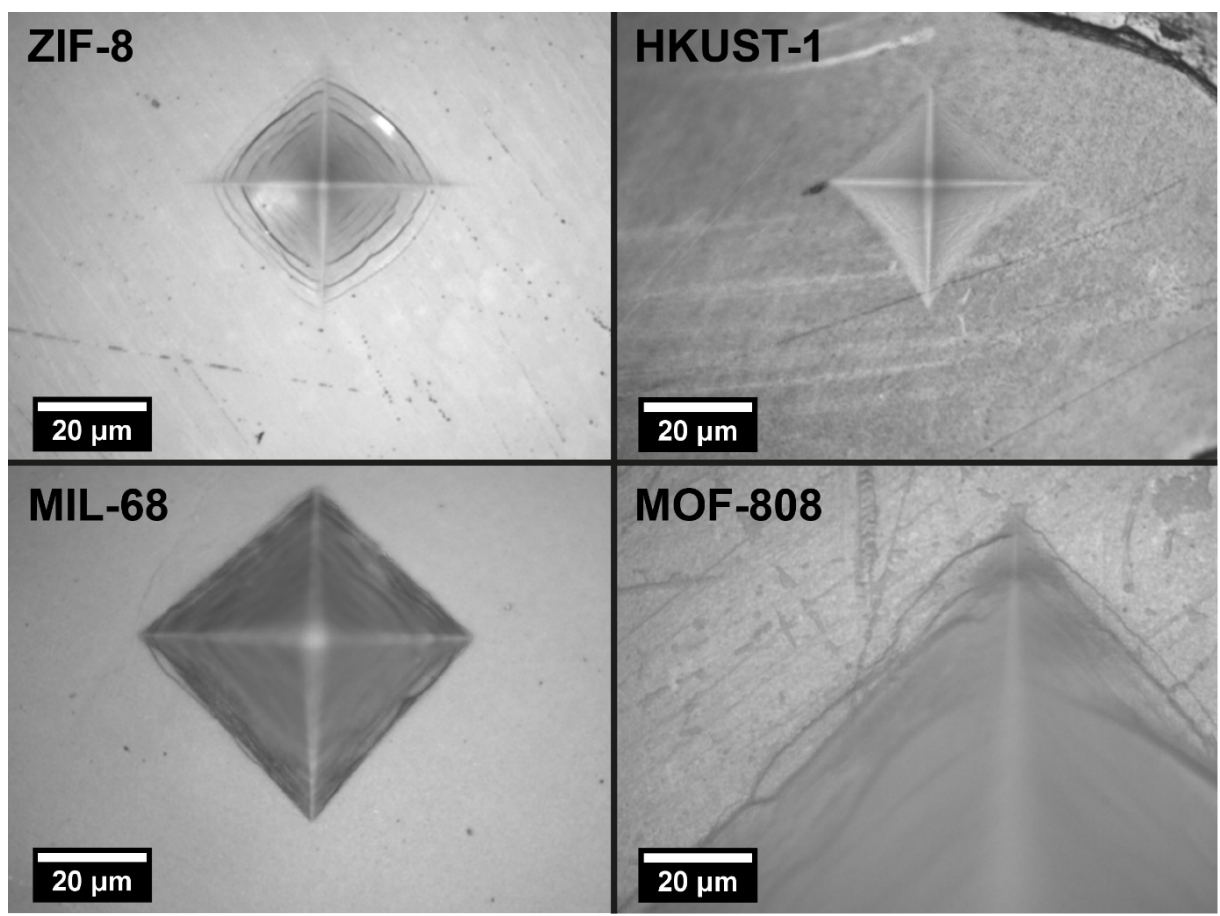

**Figure 2.** Optical micrographs of the residual indentations by Vickers microhardness, subjected to HV0.05 (50 gf ~ 0.49 N) on the four monoliths. Shear faults inside the contact area are visible. No radial cracks are detected on the sample surface.

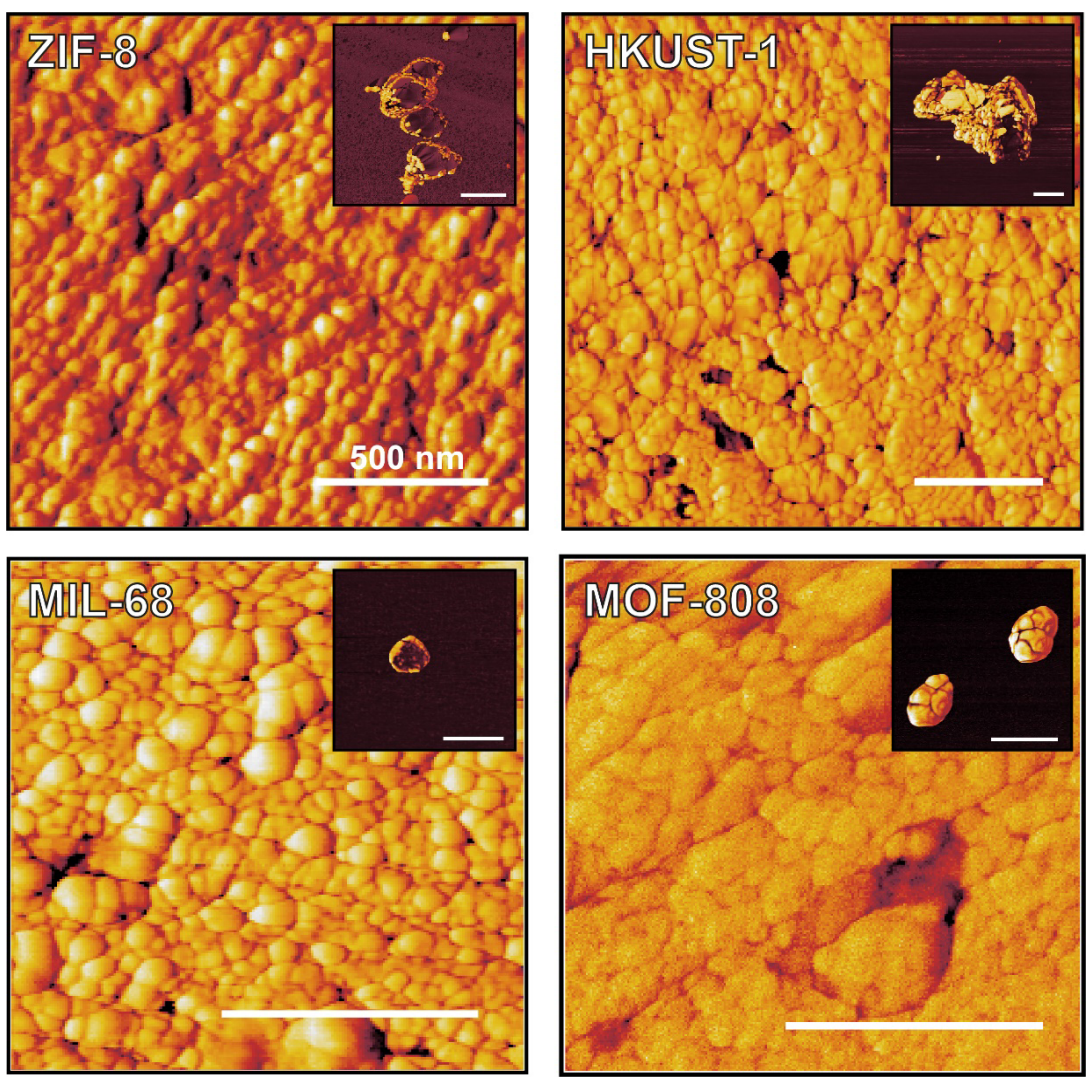

**Figure 3.** AFM phase images revealing the nanograined structure of the four monoliths. The insets show the nanostructured morphology and size of the nanocrystals constituting such a monolithic material. Two different types of aggregates can be distinguished: stack of nanoplates (ZIF-8 and HKUST-1), and lumps of nanocrystals (MIL-68 and MOF-808). Scalebar is 500 nm.

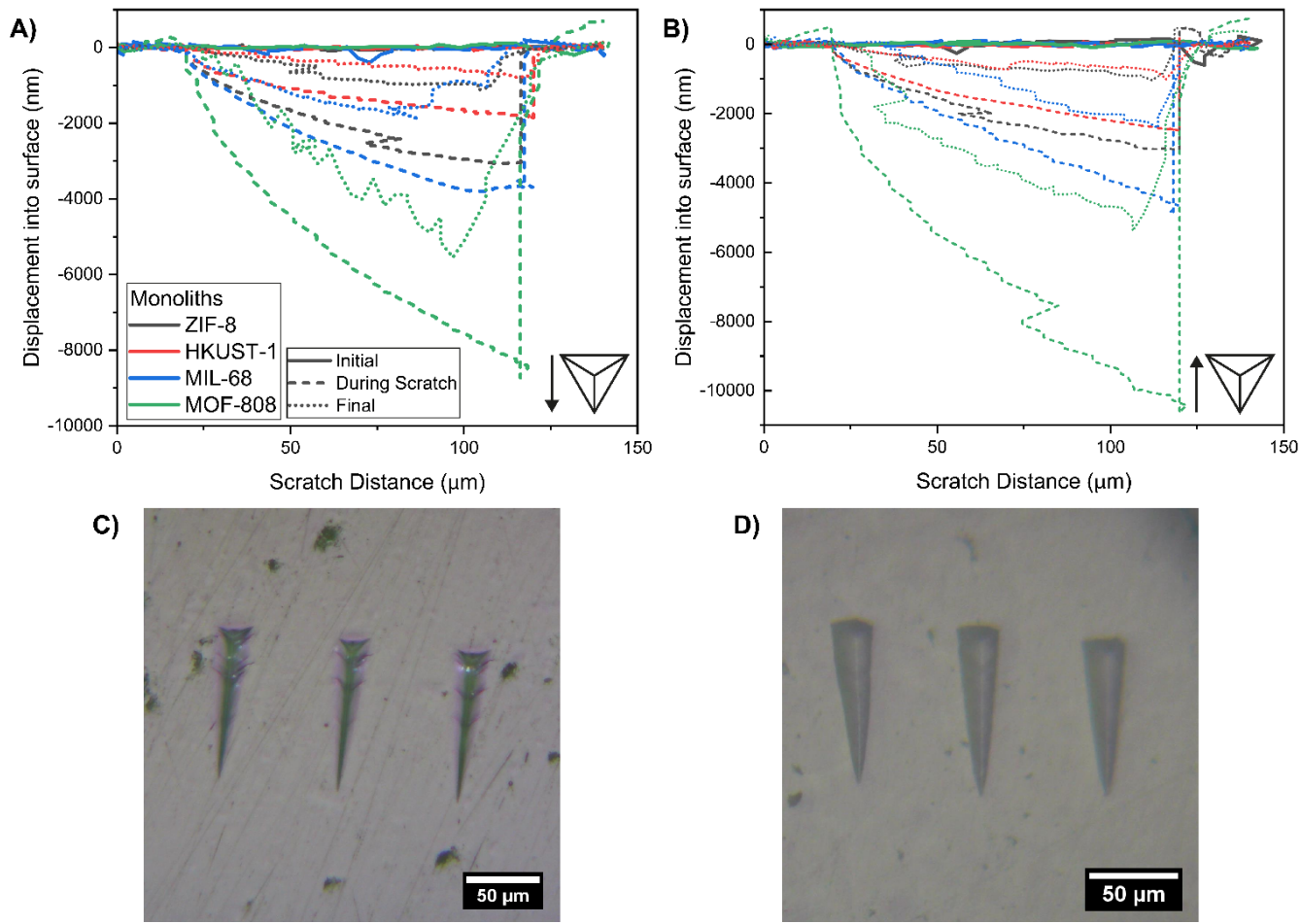

**Figure 4.** Nanoscratch profiles of the monoliths tested by ploughing mode (A) and pushing mode (B), with the displacement into the surface plotted as a function of the scratch distance. The normal load was linearly increased from 0 to 50 mN, over a total scratch length of 100 μm. Optical micrographs of the scratches on ZIF-8 (C) and MIL-68 (D) monoliths in pushing mode reveal different responses: ZIF-8 exhibits cracks formation outward from the scratching direction, while MIL-68 is visually crack-free. A summary of the optical images of all the scratches in the four monoliths in both pushing and ploughing modes is provided in Figure S12.

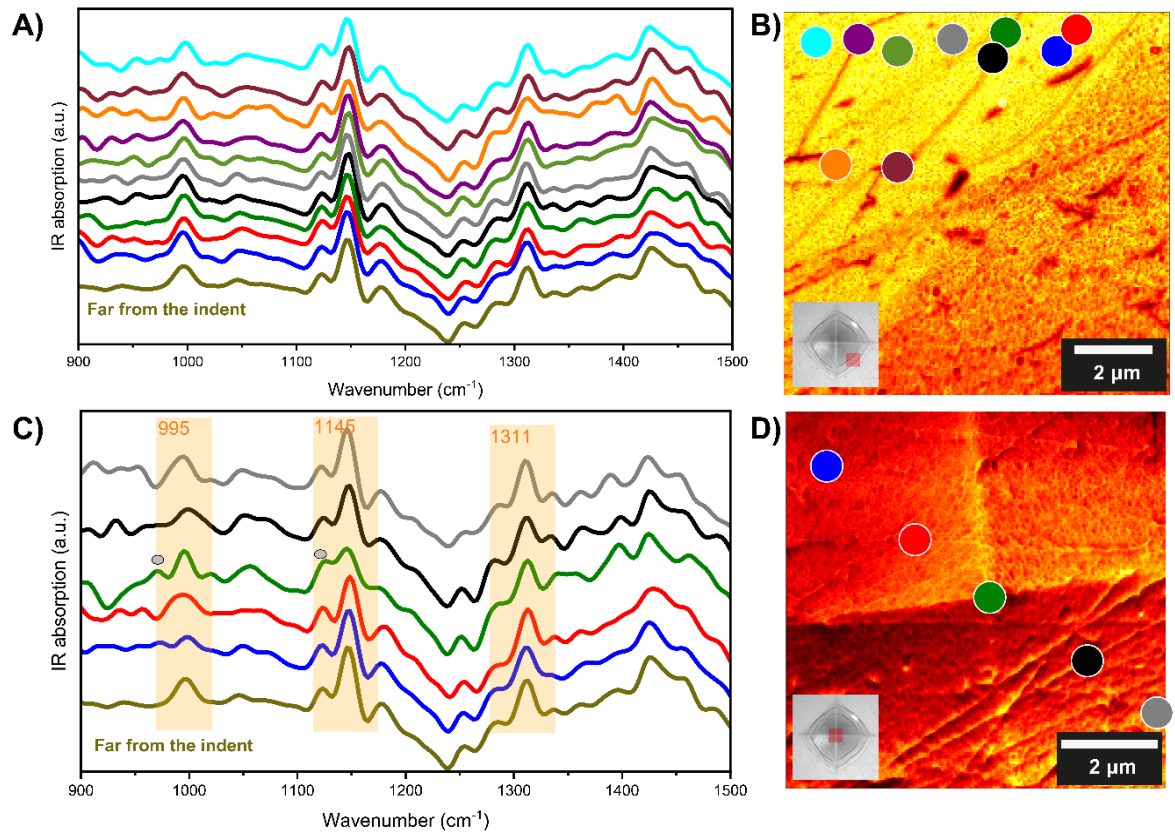

**Figure 5.** Nearfield nanoFTIR absorption spectra of ZIF-8 monolith measured locally on and around the shear faults (A) and at the indent apex inside a residual Vickers indent (C), corresponding to the positions highlighted in (B) and (D), respectively. Note that the probe size of nanoFTIR was about 20 nm. The first spectrum from the bottom in (A) and (C) was taken far away from the indented area and used as a reference for the unstrained material. The infrared absorption peaks at 995 cm$^{-1}$, and 1145 cm$^{-1}$ (highlighted by orange bands in C) are related to the characteristic vibrational modes of ZIF-8, namely, in-plane stretching of the mIm ring and C-H rocking of mIm, respectively.

# Supporting Information for

# Fracture behavior of MOF monoliths revealed by nanoindentation and nanoscratch

Michele Tricarico and Jin-Chong Tan[*]

*Email: jin-chong.tan@eng.ox.ac.uk

**This PDF file includes:**

    Figures S1 to S12
    Tables S1 to S2
    SI References

**Table S1.** Mechanical properties of the monoliths obtained from nanoindentation with a Berkovich tip. The fracture toughness was estimated by measuring the length of the radial cracks induced by nanoindentation with a cube corner tip. The values for $E^*$ (letting $v_s = 0$) and $H$ were computed by averaging the CSM data between 500 and 1000 nm. For the materials whose Poisson's ratio ($v$) is known from simulations study, also the Young's modulus was computed (see Methods). The mean and standard deviations were calculated from 32 individual indents. The residual depth at the end of the indentation test was determined from the load-depth curves (Figure S2).

| Monolith Sample | Density (g/cm³) | Indentation Modulus, $E^*$ (GPa) | Poisson's Ratio, $v$ | Young's Modulus, $E$ (GPa) | Hardness, $H$ (MPa) | Elastic energy $W_{elastic}/W_{total}$ (%) | Fracture toughness, $K_{IC}$ (MPa√m) | Residual depth – load-depth curves (nm) | Residual depth – AFM (nm) |
|---|---|---|---|---|---|---|---|---|---|
| ZIF-8 ref. (1) | 1.246 ± 0.010 | 3.78 ± 0.44 | 0.43 (2) | 3.18 ± 0.04 | 452 ± 20 | 64.1 ± 1.8 | 0.074 ± 0.023 | - | - |
| ZIF-8 This work | 1.258 ± 0.022 | 4.03 ± 0.03 | 0.43 (2) | 3.28 ± 0.03 | 534 ± 6 | 65.3 ± 1.4 | 0.081 ± 0.011 | 336 ± 5 | 246 ± 4 |
| HKUST-1 | 1.529 ± 0.043 | 15.25 ± 0.61 | 0.45 (3) | 12.16 ± 0.48 | 761 ± 53 | 39.0 ± 1.8 | 0.80 ± 0.45 | 602 ± 18 | 200 ± 1 |
| MIL-68 | 1.475 ± 0.007 | 13.24 ± 0.52 (4) | - | - | 402 ± 13 (4) | 19.2 ± 1.0 | - | 861 ± 5 | 567 ± 20 |
| MOF-808 | 1.522 ± 0.104 | 4.61 ± 0.32 | - | - | 122 ± 14 | 13.6 ± 1.1 | - | 948 ± 13 | 538 ± 5 |

| Sample | Scratch Critical Depth (nm) | Cross Profile Max Depth (nm) |
|---|---|---|
| ZIF-8 plough | 3088 ± 94 | 580 ± 72 |
| HKUST-1 plough | 1764 ± 60 | 671 ± 441 |
| MIL-68 plough | 4149 ± 339 | 1187 ± 209 |
| MOF-808 plough | 9255 ± 573 | 3266 ± 452 |
| ZIF-8 push | 3002 ± 170 | 911 ± 288 |
| HKUST-1 push | 2379 ± 138 | 502 ± 106 |
| MIL-68 push | 4895 ± 49 | 2050 ± 152 |
| MOF-808 push | 10549 ± 467 | 3762 ± 581 |

**Table S2.** Results of the nanoscratch tests on the monoliths in ploughing and pushing modes.

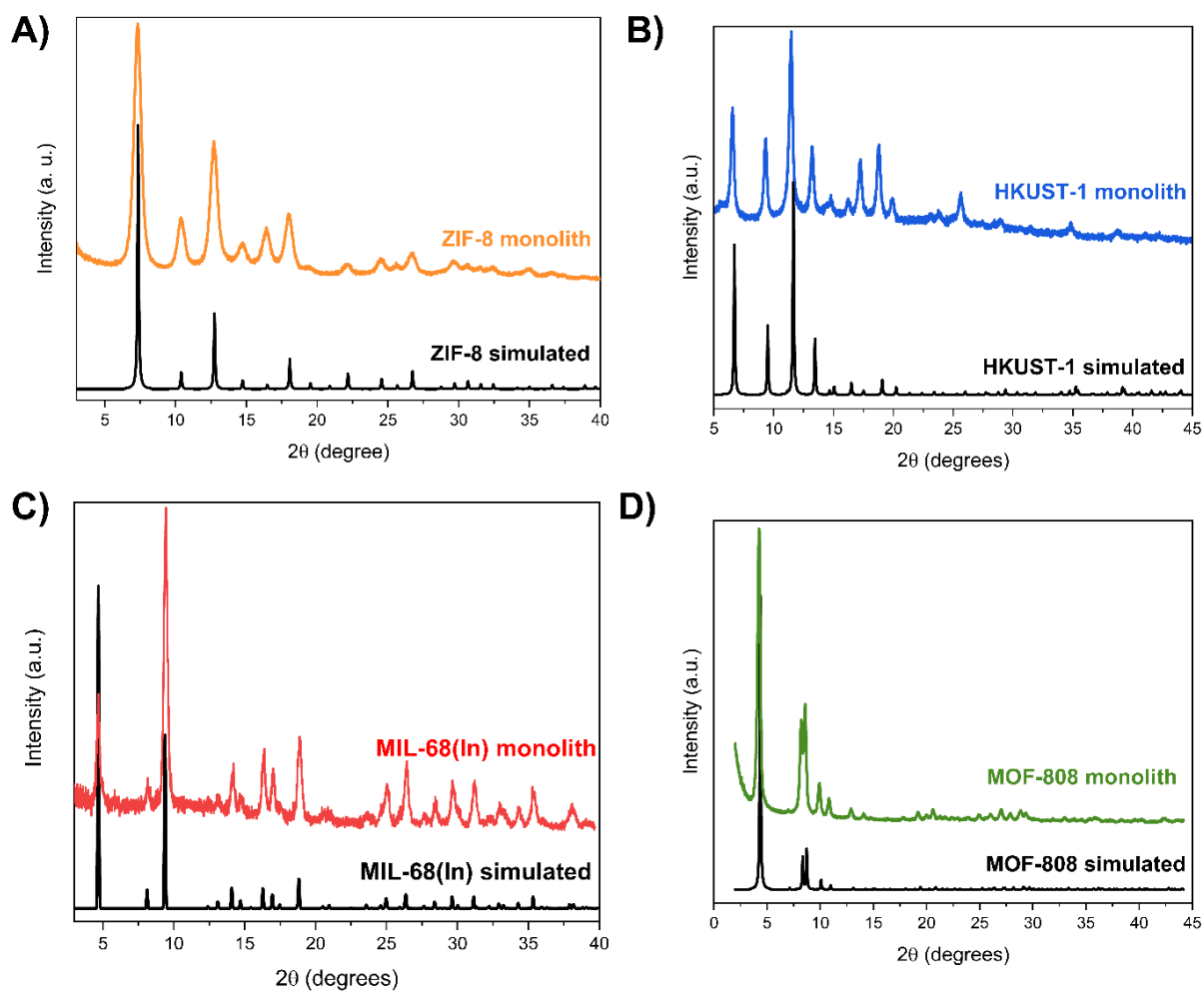

**Figure S1.** XRD patterns of A) ZIF-8, B) HKUST-1, C) MIL-68 and D) MOF-808 monoliths in comparison with the simulated patterns from the Cambridge Structural Database (CSD). CCDC codes: TUDKEJ (ZIF-8), FIQCEN (HKUST-1), LOQLEJ (MIL-68(In)), BOHWUS (MOF-808).

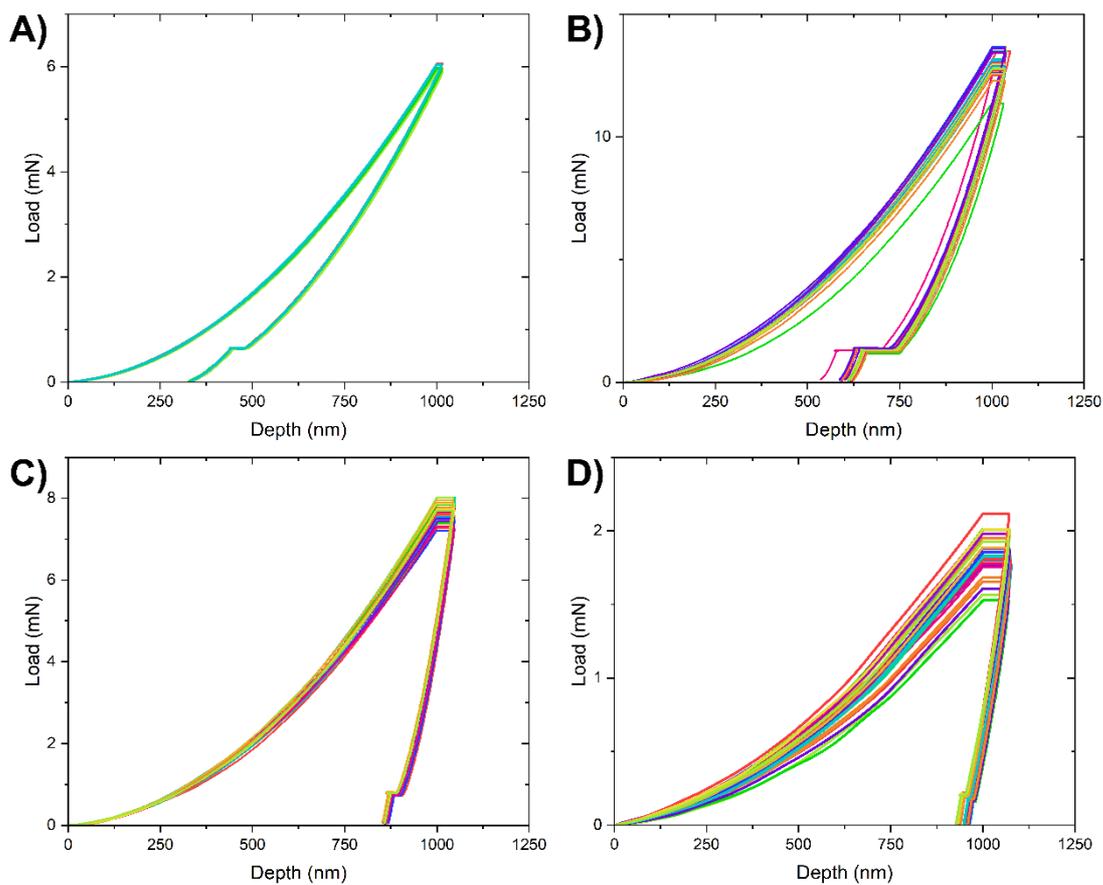

**Figure S2.** Nanoidentation load-depth curves with a Berkovich indenter to a maximum surface penetration depth of 1000 nm. (A) ZIF-8, (B) HKUST-1, (C) MIL-68 and (D) MOF-808. 32 indentations per sample were performed.

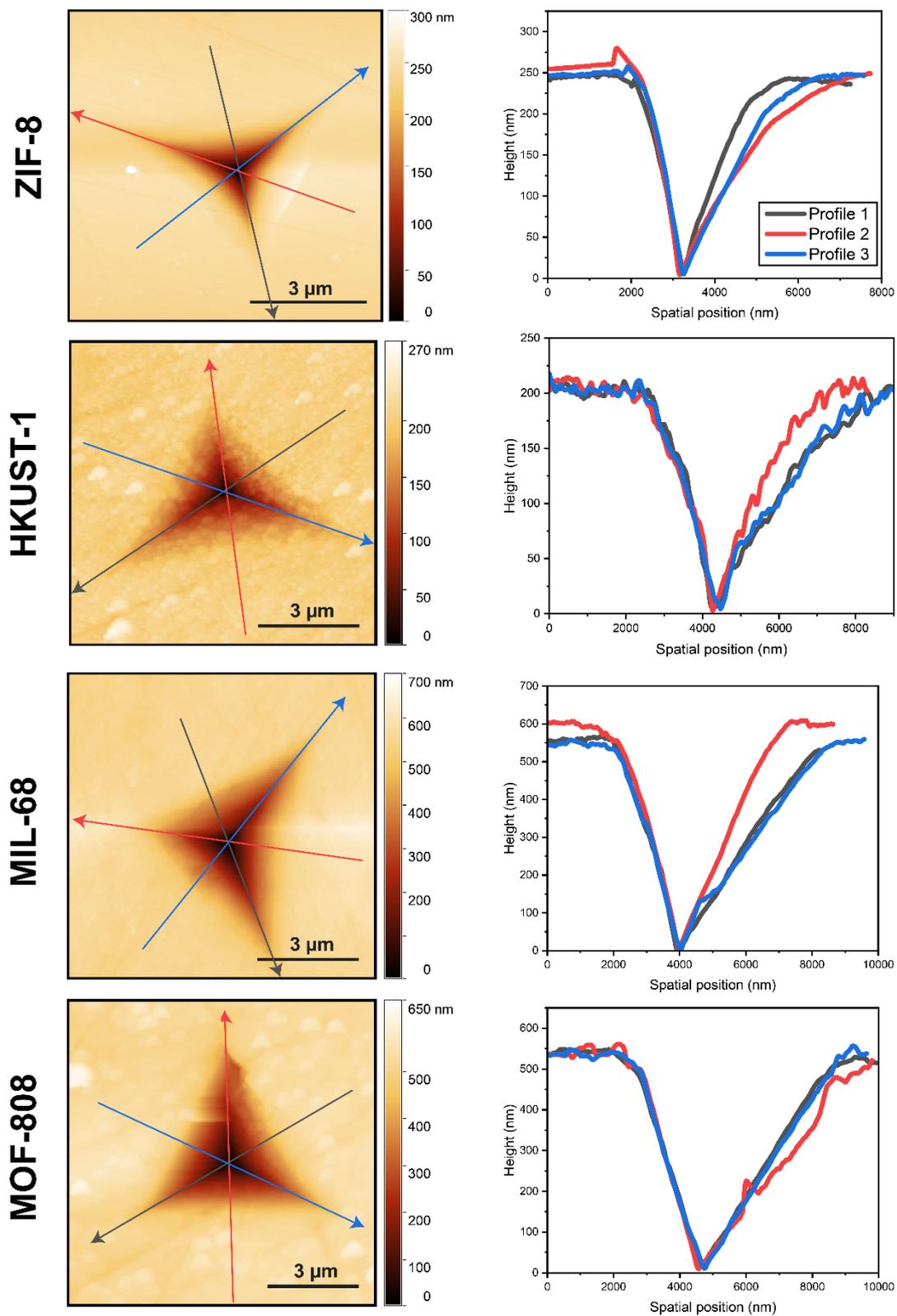

**Figure S3.** AFM topographies of the residual indents on the four monoliths. The corresponding cross-sectional profiles along the specified paths are plotted on the right.

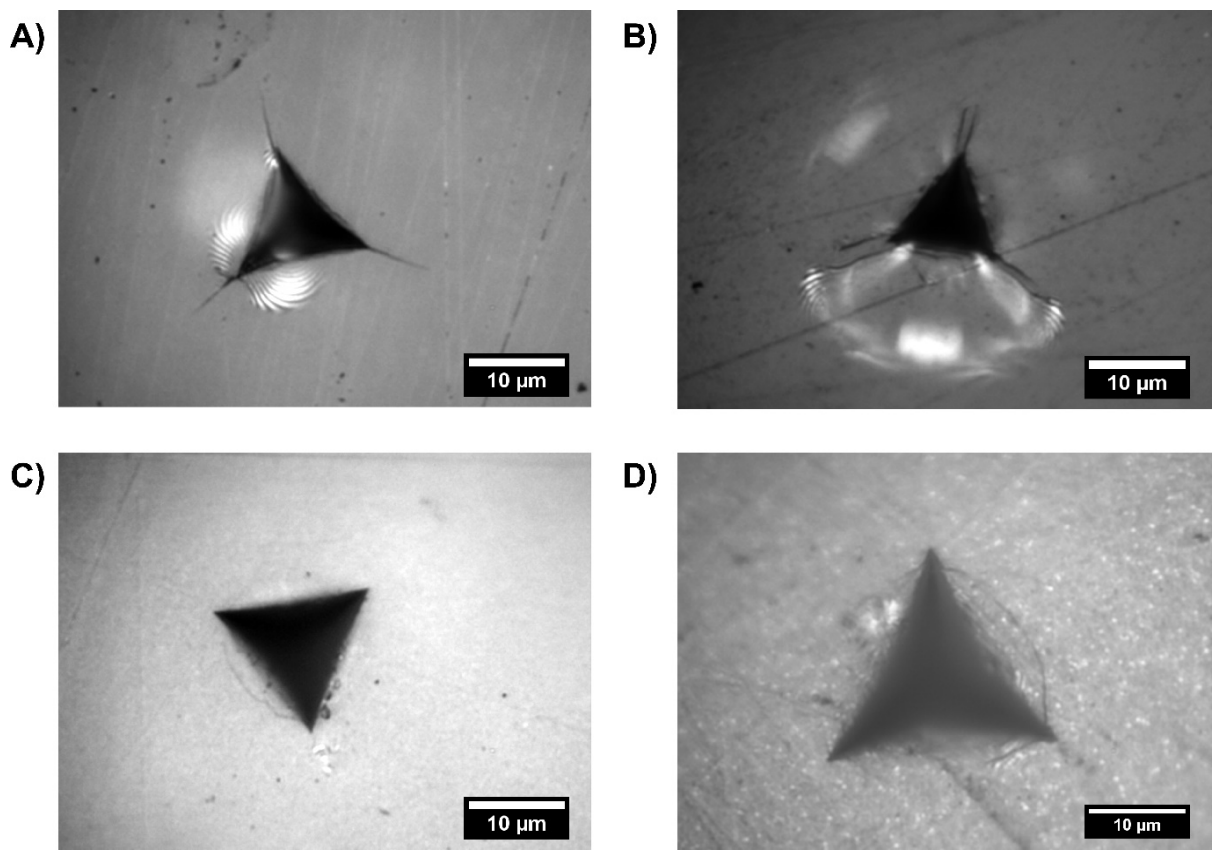

**Figure S4.** Cube corner residual indents on A) ZIF-8, B) HKUST-1, C) MIL-68, and D) MOF-808. A maximum load of 50 mN was applied in all tests.

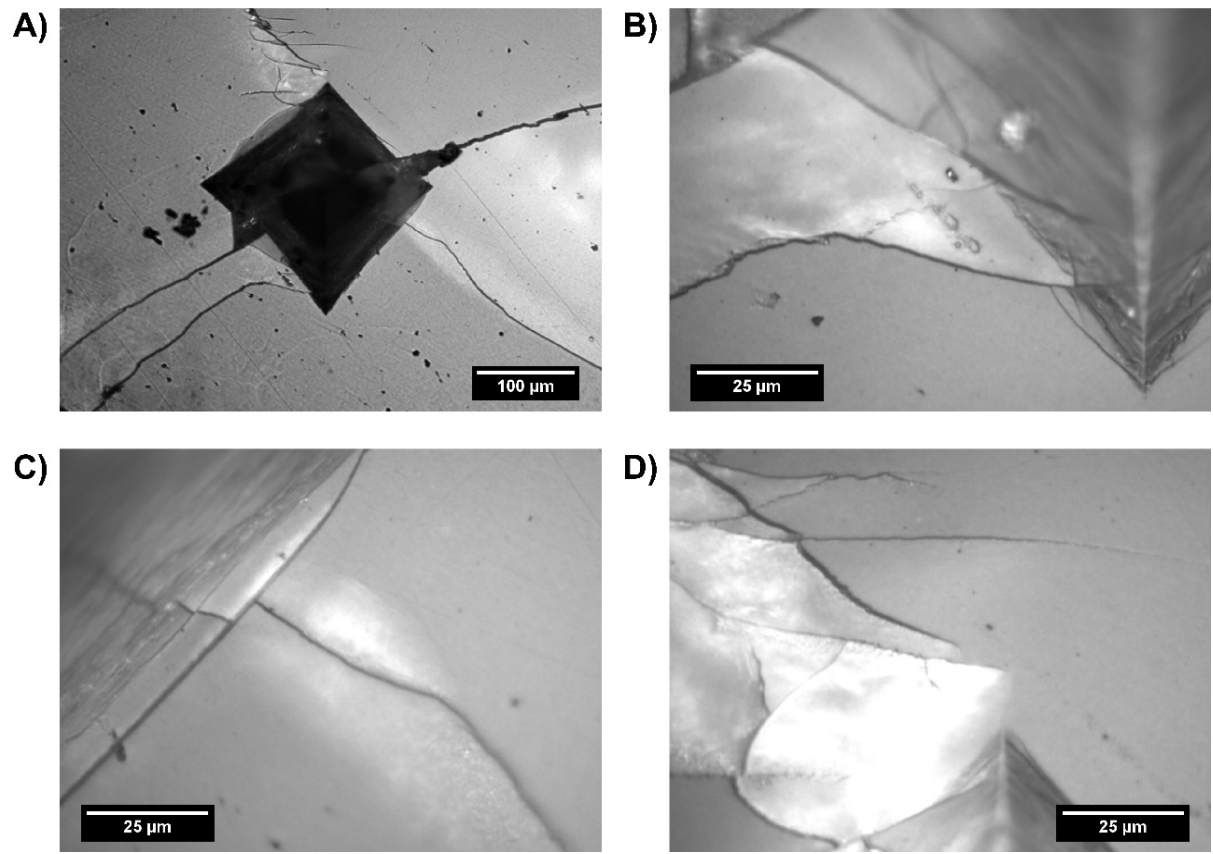

**Figure S5.** Vickers HV0.5 (~4.9 N) indent on MIL-68 monolith. Radial cracks propagate from shear faults inside the indent and deflect following low energy paths.

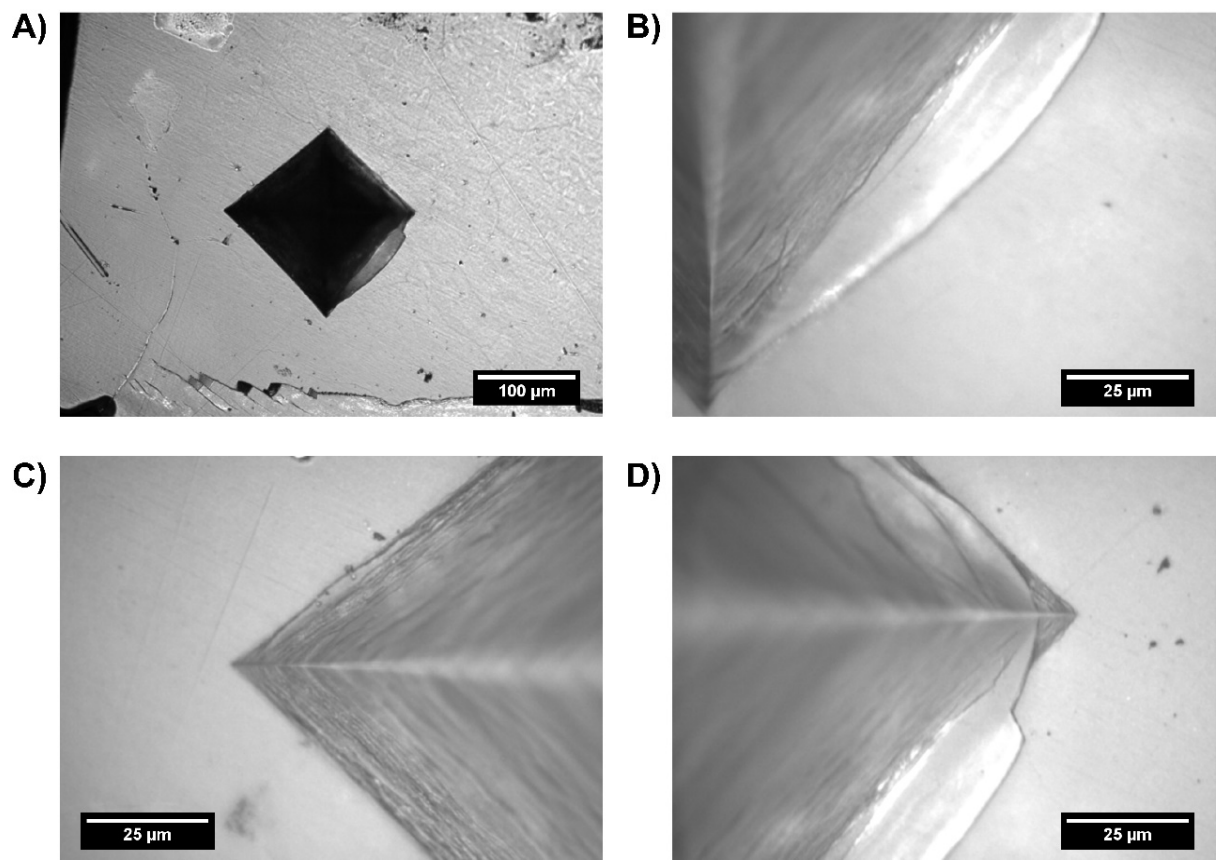

**Figure S6**. Vickers HV0.5 (~4.9 N) indent on MIL-68 monolith. In this case no radial cracks are induced.

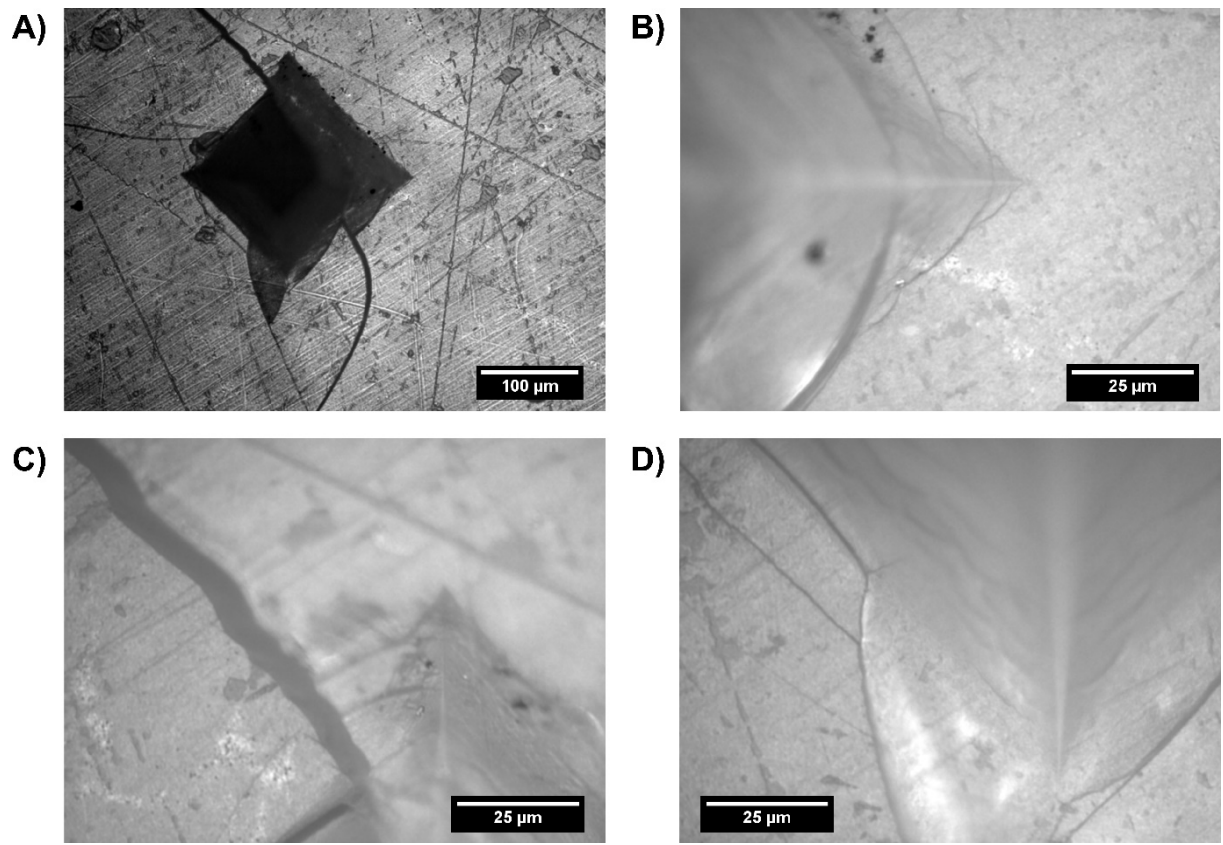

**Figure S7.** Vickers HV0.3 (~2.9 N) indent on MOF-808 monolith. Radial cracks propagate from shear faults inside the indent and deflect following low energy paths outside the indent.

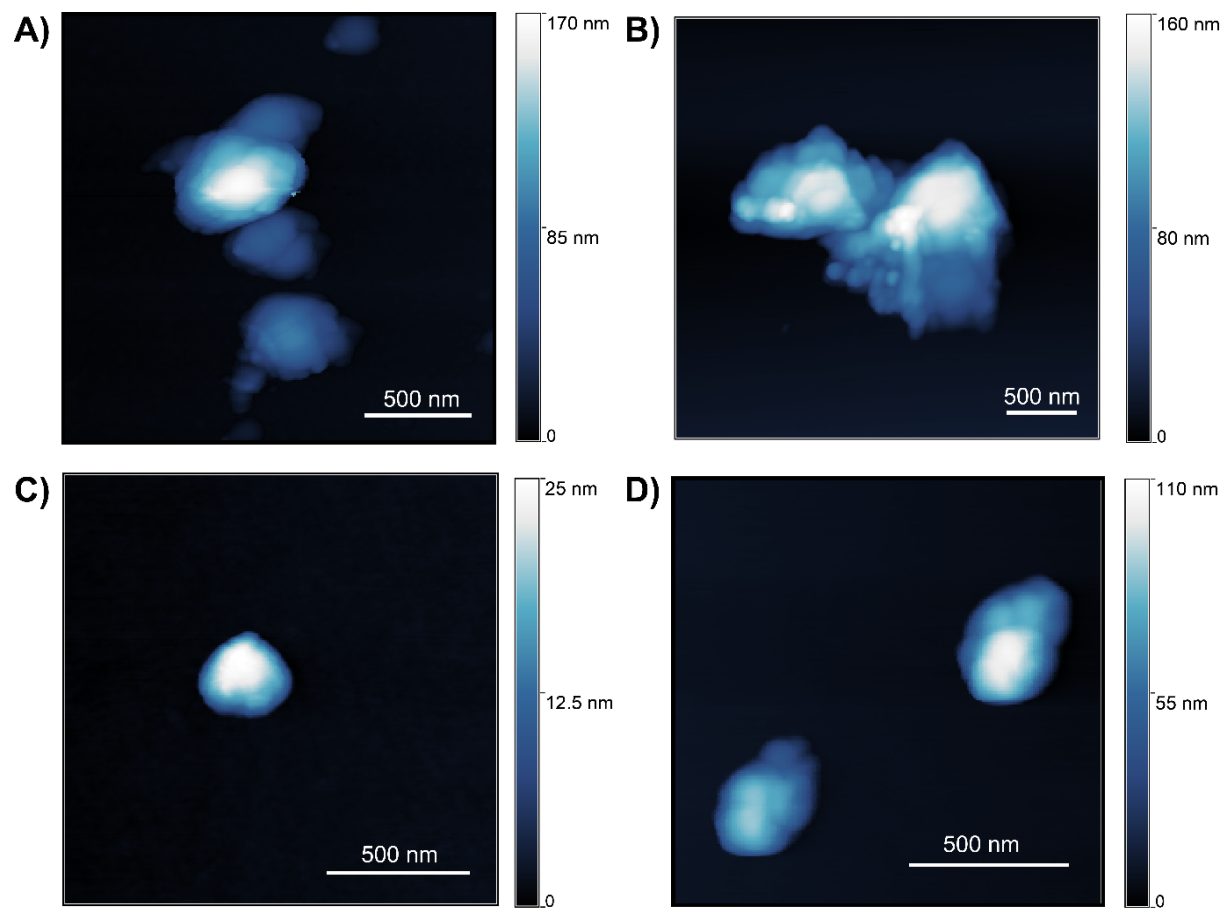

**Figure S8.** AFM height topography images revealing the morphology and size of A) ZIF-8, B) HKUST-1, C) MIL-68, and D) MOF-808 nanocrystalline aggregates.

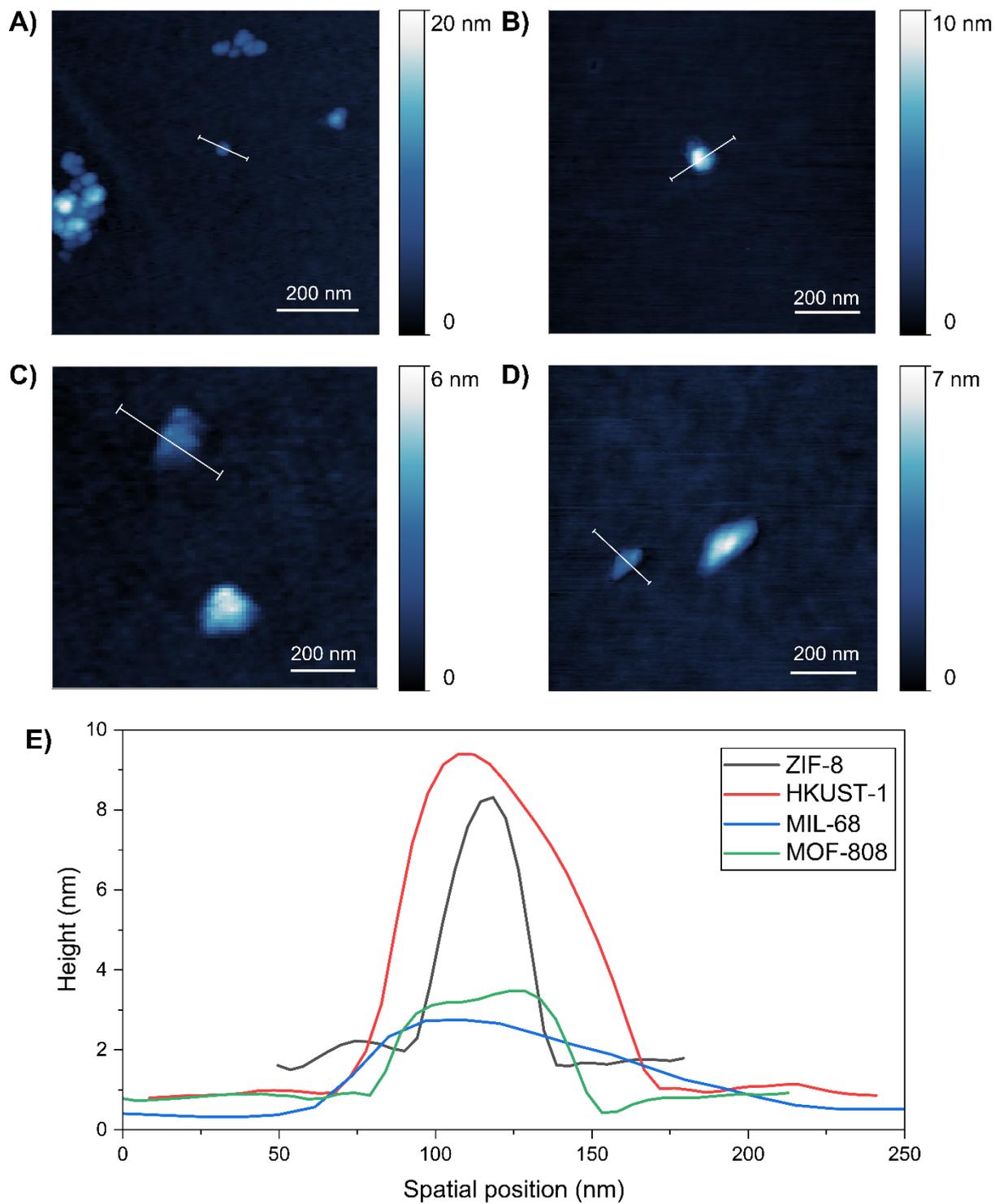

**Figure S9.** AFM height topography images revealing the nanocrystals morphology and size of A) ZIF-8, B) HKUST-1, C) MIL-68, and D) MOF-808. The corresponding line profiles are plotted in E).

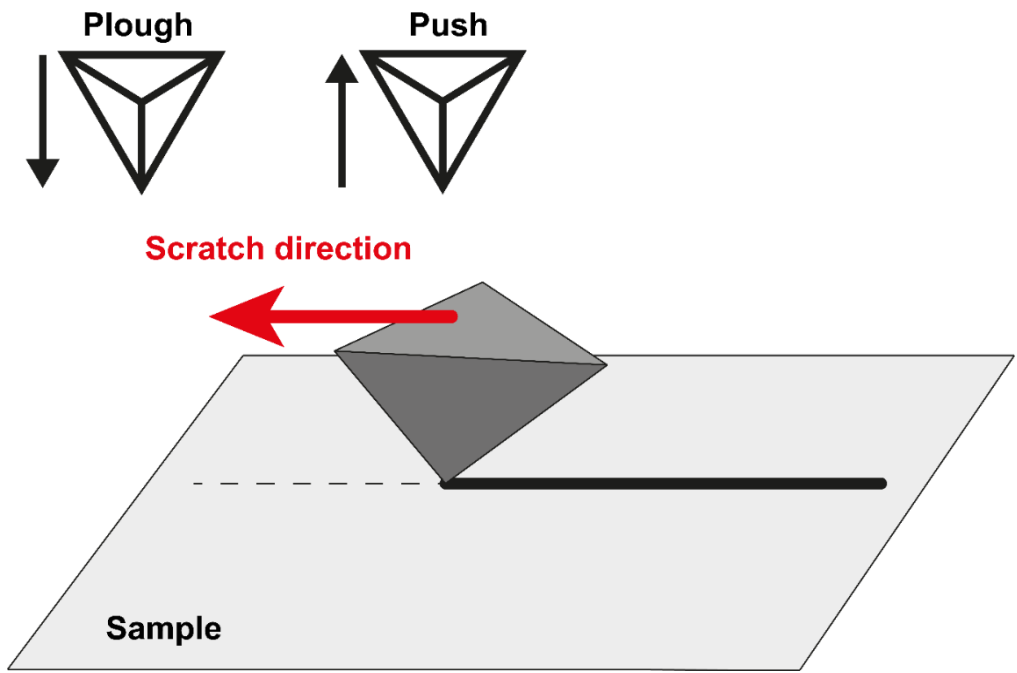

**Figure S10.** Schematic representation of a scratch test with a Berkovich tip. Since this is a three-sided pyramidal probe, two test modes are possible, depending on which end of the tip is cutting the material: ploughing (sharp end) or pushing (flat end).

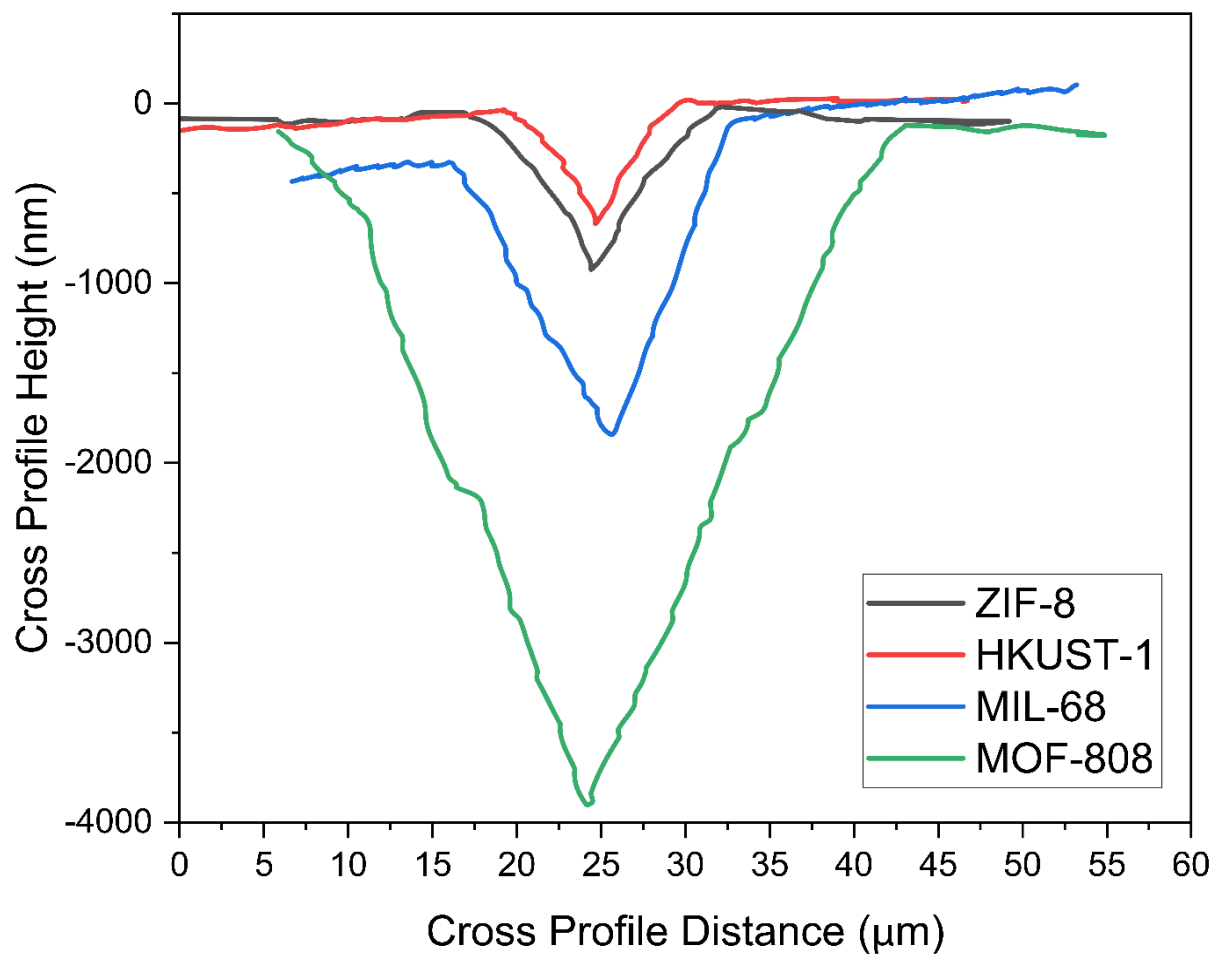

**Figure S11.** Representative mid-way cross profiles (load ~25 mN) of scratches in pushing mode on the four monoliths.

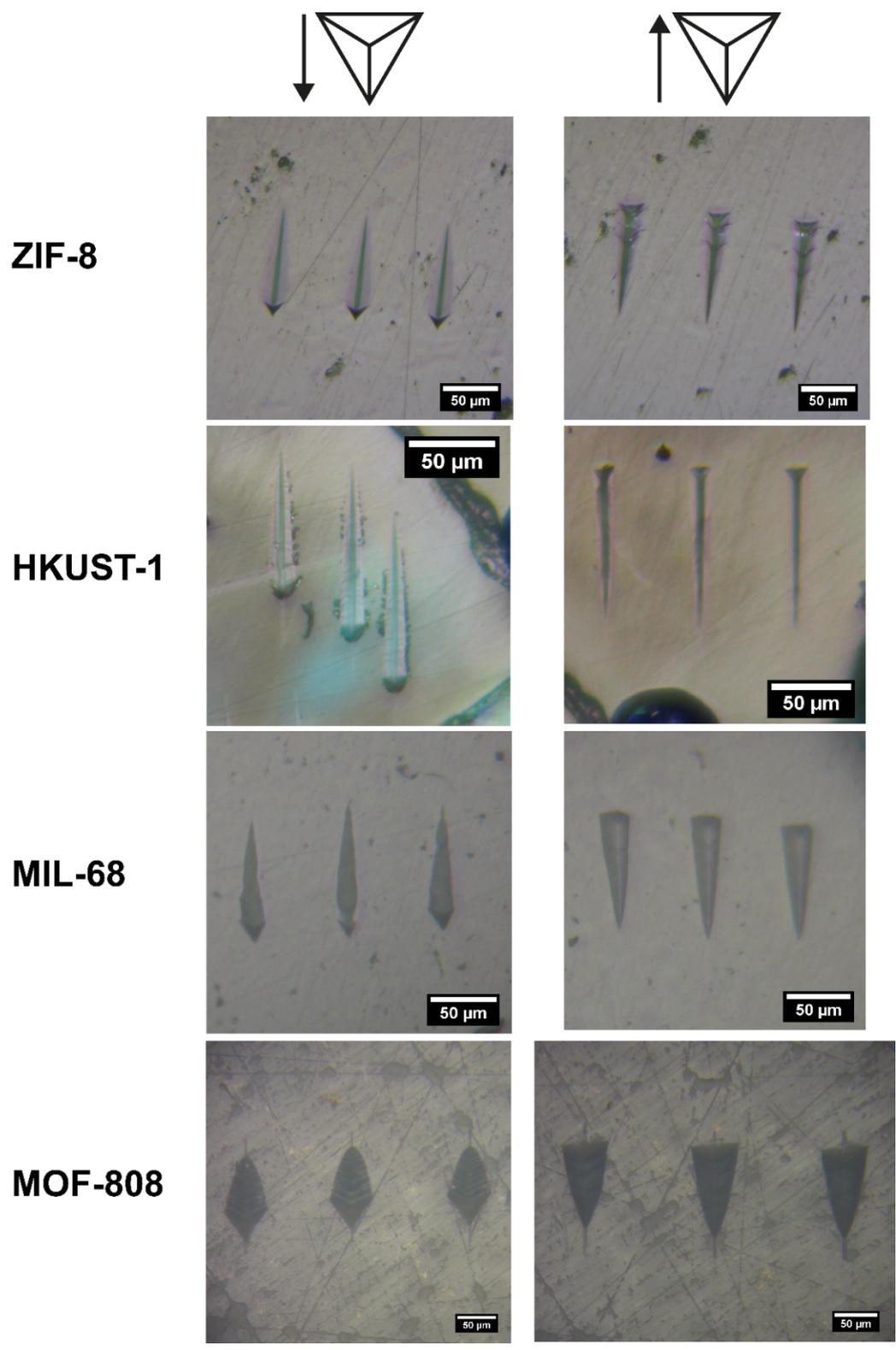

**Figure S12.** Optical micrographs of the scratches on the four monoliths in ploughing mode (left) and pushing mode (right).